\documentclass[sigconf,nonacm]{acmart}
\title[Beyond the Badge: Reproducibility Engineering as a Lifetime Skill]{Beyond the Badge: Reproducibility Engineering \\ as a Lifetime Skill}

\usepackage{url}
\usepackage{todonotes}
\usepackage[UKenglish]{babel}

\usepackage{xspace}
\usepackage{setspace}
\usepackage{graphicx}
\usepackage{textcomp}
\usepackage{xcolor}
\usepackage{wrapfig}
\usepackage{csquotes}

\usepackage[strings]{underscore}

\usepackage{tikz}
\usetikzlibrary{arrows.meta}
\usetikzlibrary{positioning,fit}
\usetikzlibrary{backgrounds}
\usetikzlibrary{decorations.pathmorphing}
\usetikzlibrary{patterns}
\usetikzlibrary{shapes}
\setcopyright{acmcopyright}
\copyrightyear{2022}
\acmYear{2022}
\acmDOI{XXXXXXX.XXXXXXX}

\acmConference[SEENG '22]{4th International Workshop on Software Engineering Education for the Next Generation}{May 17, 2022}{Pittsburgh, PA}

\acmPrice{15.00}
\acmISBN{978-1-4503-XXXX-X/18/06}

\newcommand{\webref}[2]{\textcolor{blue}{\href{#1}{#2}}} 

\newcommand{\noun}{reproducibility\xspace}

\newcommand{\eg}{\emph{e.g.}\xspace}
\newcommand{\tool}[1]{\texttt{#1}}

\begin{document}

\author{Wolfgang Mauerer}
\orcid{0000-0002-9765-8313}
\affiliation{%
  \institution{Technical University of Applied Science Regensburg}
  \institution{Siemens AG, Corporate Research}
    \country{Germany}
}
\email{wolfgang.mauerer@othr.de}

\author{Stefan Klessinger}
\affiliation{%
  \institution{Chair of Scalable Database Systems}
  \institution{University of Passau}
  \country{Germany}
}
\email{stefan.klessinger@uni-passau.de}

\author{Stefanie Scherzinger}
\affiliation{%
  \institution{Chair of Scalable Database Systems}
  \institution{University of Passau}
  \country{Germany}
}
\email{stefanie.scherzinger@uni-passau.de}

\renewcommand{\shortauthors}{Mauerer, Klessinger and Scherzinger}

\begin{abstract}
Ascertaining reproducibility of scientific experiments is receiving
increased attention across disciplines. We argue that the necessary
skills are important \emph{beyond} pure scientific utility, and that they
should be taught as part of software engineering (SWE) education.
They serve a dual purpose: Apart from acquiring the coveted badges assigned to
reproducible research,
reproducibility engineering is a \emph{lifetime skill}
for a professional industrial career in computer science.

SWE curricula seem an ideal fit for conveying such capabilities, yet they require some
extensions, especially given that even at flagship conferences like ICSE,
only slightly more than one-third of the technical papers (at the 2021 edition)
receive recognition for artefact reusability. Knowledge and capabilities in setting up
engineering environments that allow for reproducing artefacts and results over decades (a
standard requirement in many traditional engineering disciplines), writing semi-literate
commit messages that document crucial steps of a decision-making process and that are tightly
coupled with code, or sustainably taming dynamic, quickly changing software dependencies,
to name a few: They all contribute to solving the scientific reproducibility crisis,
\emph{and} enable software engineers to build sustainable, long-term maintainable,
software-intensive, industrial systems. We propose to teach these skills
at the undergraduate level, \emph{on par} with traditional SWE topics.
\end{abstract}



\maketitle

\section{Introduction}
\label{sec:intro}

Since software engineering involves complex software stacks that non-trivially interact with hardware, sharing experimental setups is anything but trivial.
Over the last decade, reproducibility of experimental results has become
recognised as a prime aspect of computer science~(CS) research. Several high-profile conferences now
award badges when results can be independently verified.

Undoubtedly, reproducibility engineering (RepEng) has become a crucial skill that today's generation of PhD students has to master.
In this position paper, we argue that these skills should already be taught (and
practised) at the undergraduate level, and we therefore designed and
conducted a course for computer science Bachelor students close to graduation.
Even when students pursue an industry career, they will greatly benefit from
recognising threats to reproducibility, how to tackle them, and how to build
long-term reproducible code.
In short, it is our conviction that students skilled in RepEng possess skills that proficient software engineers need to master (anyway).

Accordingly,
we propose a multi-faceted syllabus\footnote{We have implemented the outlined ideas in an online course, taught in the winter term
2021/22, to undergraduate students at two universities. The lecture videos
are available   \webref{https://youtube.com/playlist?list=PLbGy1_nazP3nb_ECS1qjp16mpldkdZC4k}{online} on YouTube (link in PDF).
} for teaching reproducibility engineering, and what we consider crucial skills. This includes best
practices in computer science research and industry, such as packaging entire system software stacks for dissemination.
For \emph{long-term} \noun over decades (ideally, forever), we discuss why
open source technologies (as massively employed in industry) are preferable to
approaches crafted for research.

\paragraph{Structure.}
We recap essentials on building reproduction packages.
We propose a high-level syllabus, covering social and technical best practices, as well as specific tools and technologies that are well-adopted in industry. We then discuss the literature material available for academic teaching, and conclude.

\section{Preliminaries}
\label{sec:prelims}

\paragraph{Terminology.}
Reproducibility is a cross-cutting theme
and
there are guidelines by the National Science Foundation (NSF)~\cite{NAS:2019},
    the Association of Computing Machinery (ACM)~\cite{ACMCriteria}, and the Institute of
    Electrical and Electronics Engineers (IEEE)~\cite{IEEECriteria}.
Concepts like reproducibility and replicability receive different
    interpretation, depending on the community. Even large professional associations
    like the ACM had to revise their definitions of the
    terms because of prior confusion.
   Despite their obvious relevancy, the concepts are not yet reflected in the ACM Computing Classification System\footnote{Available \webref{https://dl.acm.org/ccs}{online} (link in PDF), last updated 2012.}.

 Throughout this article, we follow the ACM terminology~\cite{ACMCriteria} (version~1.1), and regard an experiment as \emph{repeatable}, when the same team with the same experimental setup can confirm the results. An experiment is \emph{reproducible}, if it is a different team, but the same setup, that confirms the results.  Finally, an experiment is \emph{replicable}, when a different team, with a different setup, confirms the result.

\paragraph{Reproduction Packages.}
    Building a \emph{reproduction package} goes beyond providing a document object identifier (DOI) to some
repository hosting data, code, and setup instructions.
Rather, a gold-standard reproduction package~\cite{Heil2021} bundles all research artefacts
required to conduct the experiment (such as source code, libraries, or input data), and
contains a dispatcher script that allows for executing and evaluating the experiment via a \emph{single}
command.

\begin{figure}[tbh]
    \centering\begin{tikzpicture}[remember picture,
    system/.style={draw,thin,inner sep=0pt,rectangle,color=black,fill=lightgray!20},
    group/.style={draw,thin,inner sep=0pt,rectangle,color=black,fill=white!20},
    numbers/.style={draw,thin,circle,color=black,fill=yellow!20,inner sep=2pt},
    empty/.style={draw,thin,circle,color=white,fill=white!20},
    label/.style={},
    arrow/.style={-{Stealth[width=4mm,length=3mm]},black,line width=1pt},
    double arrow/.style={-{Stealth[width=4mm,length=3mm]},double,double distance=2pt,line width=1pt},
    small arrow/.style={-{Stealth[width=2mm,length=2mm]}},
]
\draw node[system,minimum width=2.3cm, minimum height=1.0cm, align=center] (docker) {Container \\ (source)};

\node[system,above=0.6cm of docker,minimum width=2.3cm, minimum height=0.5cm,align=center] (git) {Public Git Repo};
\node[system,right= 0.8cm of git,minimum width=2.3cm, minimum height=0.5cm,align=center] (patches) {Patch Stack};
\node[system,left= 0.8cm of git,minimum width=2.3cm, minimum height=0.5cm] (binaries) {Binaries};

\node[system,left = 1.6cm of docker,minimum width=1.5cm, minimum height=1cm, align=center] (recipe) {Build \\ Recipe};

\node[system,right= 1.6cm of docker,minimum width=1.5cm, minimum height=1cm, align=center] (dockerBinary) {Container\\(binary)};

\node[group, below = 0.7cm of docker,minimum width=7.3cm, minimum height=1.2cm] (tarball) {};
\node[label,anchor=north west,align=left, inner sep=5pt] at (tarball.north west) {Experiment Execution Package};
\node[system,anchor=south east, xshift=-5pt, yshift=5pt, minimum width=2.6cm, minimum height=0.5cm,align=left] at (tarball.south east) (queries) {Data + Generators};
\node[system, left = 0.35cm of queries,minimum width=1.8cm, minimum height=0.5cm] (dispatcher) {Dispatcher};
\node[system, left = 0.35cm of dispatcher,minimum width=1.8cm, minimum height=0.5cm] (evaluation) {Evaluation};

\draw[arrow] (binaries.south) -- ++(0.0,-0.175) -- ++(3.11,0) -- (docker.north);
\draw[arrow] (git.south) to[out=270, in=90] (docker.north);
\draw[arrow] (patches.south) -- ++(0.0,-0.175) -- ++(-3.11,0) -- (docker.north) ;

\draw[double arrow] (recipe.east) to[out=0, in=180] (docker.west);
\draw[double arrow] (docker.south) to[out=270, in=90] (tarball.north);
\draw[double arrow] (docker.east) to[out=0, in=180] (dockerBinary.west);
\draw[double arrow] (dockerBinary.south) -- ++(0,-0.3) -- ++(-3.515,0) -- (tarball.north);

\node[numbers,right=0.45cm of recipe] (recipeLabel) {\large 1 };
\node[numbers,right=0.45cm of docker] (tarballLabel) {\large 2 };
\node[numbers,above = 6pt of tarball,xshift=15pt,yshift=-0.15em] (tarballLabel) {\large 3 };

\node[label,below = 0.75cm of tarball.west,xshift=-0.5cm,text depth=0] (l1l) {A};
\node[label,right = 0.4cm of l1l,text depth=0] (l1r) {B, B integrates A};
\draw[small arrow,black,line width=1pt] (l1l.east) to[out=0, in=180] (l1r.west);

\node[label,below = 0.75cm of tarball.east, xshift=-0.8cm,text depth=0] (l2r) {B, B is produced by A};
\node[label,left = 0.4cm of l2r, anchor=east, align=right,text depth=0] (l2l) {A};
\draw[small arrow,black,double,double distance=1.3pt,line width=0.8pt] (l2l.east) to[out=0, in=180] (l2r.west);

\end{tikzpicture}\vspace*{-2.5em}
    \caption{Building a reproduction package~\protect\cite{BTW2021}.}
    \label{fig:repro}
\end{figure}

Figure~\ref{fig:repro} (adapted from~\cite{BTW2021}) shows a state-of-the-art setup.
Based on system binaries, external and internal code in
    git repositories, and patch stacks with changes to existing components,
    a build
    recipe induces generation of a host-system independent Docker container as
    (static and immutable) build environment for measurement binaries (1). Additionally, a Docker container with pre-built binaries, devoid of any external dependencies, is created (2). The Docker container creates an experiment execution package (3) that can be deployed on
    cloud systems, or on local hardware, without any dependence on
    target-system provided artefacts. The experimental runs generate
    data, which are post-processed, evaluated  and visualised by scripts in the experiment execution package.

\section{A Multi-Level Syllabus}

We argue that reproducibility engineering should find its way into undergraduate curricula, anchored in software engineering education.
By targeting a clearly scoped audience  (rather than STEM disciplines in general), we can address matters
\emph{to the point}, and provide actionable advice beyond the
mechanical use of tools, or compliance to formal processes.
In sketching out a syllabus, we propose a multi-level approach,  and distinguish social and technical best practices. We further review specific tools and technologies.

\subsection{Best practices: Social}

    A reproduction package should contain as many details as necessary, but
    must not overwhelm. Instead of
    minutiæ of how results were obtained,
    a reproduction package presents
    a concise and balanced view of the \emph{outcome} of an effort.
    While any \emph{structural decisions} are worth
    preserving, the temporal order of the \emph{thought process} that led to
    intermediate results or to said decisions, is usually not.

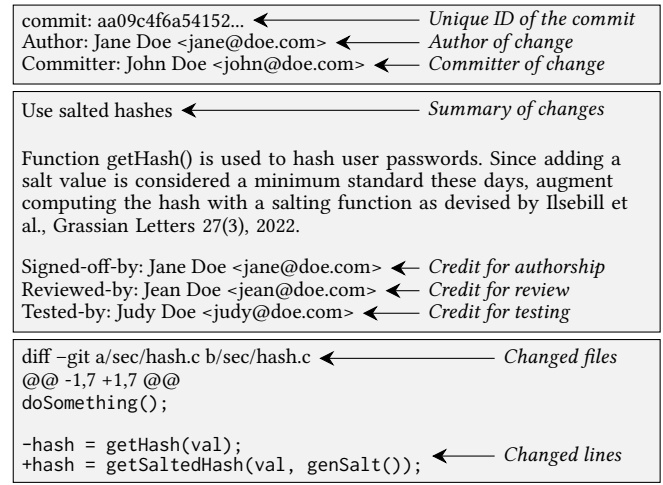
\begin{figure}[htb]
    \begin{tikzpicture}[remember picture,
annotation/.style={color=black,font=\small\itshape, align=left, anchor=west},
label/.style={font=\small, align=left, anchor=west, execute at begin node=\setlength{\baselineskip}{5pt}},
code/.style={font={\fontsize{8pt}{12}\ttfamily}, align=left, anchor=west, execute at begin node=\setlength{\baselineskip}{5pt}},
arrow/.style={-{Stealth[width=2mm,length=2mm]},black,thin},
]

\draw[fill=lightgray!20,draw=black] (0,0.2) rectangle ++(8.6,-1.05);
\draw[fill=lightgray!20,draw=black] (0,-0.95) rectangle ++(8.6,-3.2);
\draw[fill=lightgray!20,draw=black] (0,-4.25) rectangle ++(8.6,-1.9);

\node[label] at (0.0, 0.0) (commit) {commit: aa09c4f6a54152...};
\node[annotation] at (5.4, 0.0) (commitComment) {Unique ID of the commit};
\draw[arrow] (commitComment.west) to[out=180, in=0] (commit.east);

\node[label] at (0.0, -0.3) (author) {Author: Jane Doe <jane@doe.com>};
\node[annotation] at (5.4, -0.3) (authorComment) {Author of change};
\draw[arrow] (authorComment.west) to[out=180, in=0] (author.east);

\node[label] at (0.0, -0.6) (commiter) {Committer: John Doe <john@doe.com>};
\node[annotation] at (5.4, -0.6) (commiterComment) {Committer of change};
\draw[arrow] (commiterComment.west) to[out=180, in=0] (commiter.east);

\node[label] at (0.0, -1.2) (message) {Use salted hashes};
\node[annotation] at (5.4, -1.2) (messageComment) {Summary of changes};
\draw[arrow] (messageComment.west) to[out=180, in=0] (message.east);

\node[label,text width=\linewidth-2mm] at (0.0, -2.3) (messageBody) {Function getHash() is used
to hash user passwords. Since adding a salt value is considered a minimum standard these
days, augment computing the hash with a
salting function as devised by Ilsebill et al., Grassian Letters 27(3), 2022.};
\node[label] at (0.0, -3.3) (signoff) {Signed-off-by: Jane Doe <jane@doe.com>};
\node[annotation] at (5.4, -3.3) (signoffComment) {Credit for authorship};
\draw[arrow] (signoffComment.west) to[out=180, in=0] (signoff.east);

\node[label] at (0.0, -3.6) (review) {Reviewed-by: Jean Doe <jean@doe.com>};
\node[annotation] at (5.4, -3.6) (reviewComment) {Credit for review};
\draw[arrow] (reviewComment.west) to[out=180, in=0] (review.east);

\node[label] at (0.0, -3.9) (test) {Tested-by: Judy Doe <judy@doe.com>};
\node[annotation] at (5.4, -3.9) (testComment) {Credit for testing};
\draw[arrow] (testComment.west) to[out=180, in=0] (test.east);

\node[label] at (0.0, -4.5) (changedFiles) {diff --git a/sec/hash.c b/sec/hash.c};
\node[annotation] at (6.4, -4.5) (changedFilesComment) {Changed files};
\draw[arrow] (changedFilesComment.west) to[out=180, in=0] (changedFiles.east);

\node[label] at (0.0, -4.8) (diff) {@@ -1,7 +1,7 @@};
\node[code] at (0.0, -5.1) {doSomething();};
\node[code] at (0.0, -5.8) (diffTwo) {-hash = getHash(val); \\
+hash = getSaltedHash(val, genSalt());};
\node[annotation] at (6.4, -5.8) (diffTwoComment) {Changed lines};
\draw[arrow] (diffTwoComment.west) to[out=180, in=0] (diffTwo.east);

\end{tikzpicture}\vspace*{-2em}
    \caption{A technical software change accompanied by a trail of
    responsibility, revealing the rationale behind the change.}
    \label{fig:commit}
\end{figure}

Industry has established conventions~\cite{Ramsauer:2019} on documenting
software changes (at the granularity of individual commits)
to provide an understanding of the evolution of large
software systems. These conventions can also be applied to documenting research progress.
Such \emph{trails of responsibility} (which persons
authored changes together, who provided reviews, who participated
in design decisions, etc.) are routinely created outside academia
(contrariwise to the care taken in giving credit and attribution in
published papers, this approach is not established in many areas of
computer science). Figure~\ref{fig:commit} shows an example: It contains
the technical change in form of a diff (bottom part), and metadata
(unique hash, author and committer) as they are provided by version
control systems like git. Apart from this information, as it is widely
used in repository mining research~\cite{zimmermann:2005}, the commit also includes
a summary of the change, and a rationale (brief for the sake of example)
\emph{why} the change is necessary, and \emph{which} techniques are employed.

The commit can be seen as a form of communication with
fellow humans instead of mere instructions for machines,
following Knuth's seminal  \emph{literate programming} concept~\cite{Knuth:1984}.
To create \emph{readable histories}, we suggest to introduce the pragmatic customs
developed in large, international and multi-disciplinary infrastructure projects
(such as the Linux kernel) in software engineering courses.

\subsection{Best practices: Technical}

We need to provide actionable guidance on how to implement the suggested procedures and
approaches. This entails covering the necessary means \emph{end-to-end}, from
preparing all software components required to perform experiments, running
analysis code and evaluations, and to creating insightful visualisations.

Building research artefacts depends on external sources, whose long-term
availability is often not sufficiently considered. Particular care is taken to raise awareness
for identifying potential issues when aiming at \emph{reproducible builds}.

The \emph{granularity of packaging artefacts} is an important discussion point:
Should reproduction packages start directly with building
the operating kernel from source, to establish absolutely identical  conditions
given identical hardware, or is it sufficient to package custom code that leverages
any suitable execution platform? Likewise, should and can reproduction efforts re-compute all
derived results, or start with
data obtained from long-running calculations?

Another dimension concerns the variability
of programming language, compiler and toolchain,
and the distinction between build, execution, and evaluation platform.
Each of the combinations that appear
in practice have peculiarities worth discussing.

Furthermore, we consider the technical ramifications of different types of
reproducibility introduced in Section~\ref{sec:prelims}: Depending on what type of quantities
are handled---physical quantities like time or energy consumption,
numeric results from deterministic or stochastic processes, etc.---,
different means ensure that it is possible to decide whether a
reproduction attempt is successful.

Using closed-source, \emph{proprietary components} creates hurdles for other researchers,
and should be avoided in ideal open science. However, relying on  proprietary components cannot be completely avoided, so we need to discuss how to best
handle such scenarios.

Advanced numerical techniques that require accelerator hardware such as GPUs
receive increasing attention in machine
learning and artificial intelligence projects. The involved software stacks do
not only contain binary-only, proprietary components whose licenses place
obstacles on distribution, but also do not
play well with virtualisation and containerisation approaches. We need to discuss how to handle
these issues specific to \emph{dealing with hardware}.

Finally, we need to address how to properly package
artefacts and ensure their \emph{long-term availability}. Besides using
well-structured hierarchies and self-documenting package formats, we
address dual strategies towards short- and long-term reproducibility:
The latter aims at decades of reproducibility, at a higher cost to the
reproducers, while the former accepts technologies and platforms that
are not certified for \emph{DOI-safety}, but allow for easier integration into
standard development workflows. This balances advantages of long-term
reproducibility with the ease of continuous development.

\subsection{Tools and Technologies}
We need to demonstrate tools that implement the previously discussed techniques.
Primarily, we focus on Linux/\hspace*{0mm}Unix-based com\-mand-line tools,
as these are also conveniently available on standard operating systems
(Windows/Mac OS). This does not necessarily hold the other way
(\eg, Powershell), and for GUI approaches. A
small subset of the tool functionality is sufficient for reproducibility
engineering, and command-line based approaches are helpful locally and
on servers.
We suggest starting with means for easy, but effective, \emph{low-threshold automation}
based on efficient interaction with shells, pipelined processing of data, and
\emph{glue languages} such as python, R or Matlab.

\emph{Non-linear history rewriting} provided by git
allows for transforming chronological records into a readable, consistent
research process documentation
by splitting, merging, and re-ordering. Outside of software
engineering, we have encountered little knowledge of such transformations,
yet they are crucial to ascertain long-term human understandability.

\emph{Virtualisation and containers}~\cite{Boettiger:2015} play a major role in our strategy: For one, they avoid having to deal with different versions and compositions of
compilers, libraries, and system software when building artefacts. Also, they
allow for establishing a completely self-contained environment without external
internet-based dependencies that remains operational even decades after the
original sources have vanished (figuratively and literally, research is even possible
when trapped on a remote island).
Careful engineering of containers ensures they are suitable for
reproduction tasks. The appropriate techniques and patterns should therefore be introduced.

The \tool{\webref{https://salsa.debian.org/reproducible-builds/reprotest}{reprotest}}
tool collection is a recommended means to satisfy
requirements for reproducible builds: By varying environmental
parameters like user ID, folder names, or compiler settings, the tools
detect issues that do not surface when a single researcher builds code on
the always-same machine.
Such setups lead, in our experience,  to important insights on subtle
sources of errors caused by implicit, yet common misconceptions.
While such tools are routine for developers of distributions like
Debian, and also key to long-term \webref{https://www.cip-project.org/}{industrial
maintainability} of software, we find them not yet sufficiently integrated into SWE
curricula.

We suggest to implement \emph{preparing and documenting experiments}
using \tool{\webref{https://yihui.org/knitr/}{knitr}}, which
is not unsimilar to the paradigm of literate
programming~\cite{Knuth:1984,Claerbout:2005}.
It also allows for creating self-contained papers that realise end-to-end
reproduction. \emph{Electronic notebooks} like \tool{\webref{https://jupyter.org/}{Jupyter}} are a recommended variant.

How to \emph{describe and pin down the execution environment} is a further challenge.
Typical hardware specifications in published research describe the  experimental conditions  along the lines of
\enquote{Linux version 5.1.92 on a Dull Powervortex 4711 with 24~GiB of RAM was
used}. This is insufficient for reliable reproduction---non-standard kernel extensions
that may vary widely depending on the distribution, specific settings for tuning
parameters that exist  in a wide variety on every system, and many other factors that
may easily be dismissed as irrelevant can impact measurements by orders of magnitude.
We recommend discussing means of faithfully recording the execution conditions of
computational experiments.

Students should acquire hands-on experience in \emph{reproducing experimental outcomes}.
Retracing the work of others increases awareness for (and appreciation of) high-quality reproduction packages.

\subsection{Special Cases}
While software engineering can often be decoupled
from details of the target environment (CPU architecture,
OS version, \dots), special-purpose hardware introduces
additional reproducibility requirements. We find that
general-purpose graphical processing units (GPGPUs) necessitate software stacks
that exceed standard compilers considerably in size, and introduce (a) strong
interdependencies between software component versions and (b)
dependencies on specific features that might only find intermittent
support in hardware. Both stress the need to teach implementing less
performant, but generic alternatives, and how to store
intermediate results obtained from HW accelerators for further processing.
Similar considerations hold for tensor processing units (TPUs) and other
AI accelerators. Quantum computing,  starting to receive
interest from the software engineering community~\cite{qsa,zhao2020quantum},  additionally needs to deal with globally unique hardware semi-prototypes~\cite{q-saner}.

\section{Teaching Material}

\paragraph{Textbooks.}
While
a number of books on the subject itself have been published, they are either (a) edited collections
of articles written by different authors and lack a central \emph{leitmotif}, (b) focus on
high-level aspects of reproducibility, or (c) consider very narrow domains.
For instance, recent books discuss reproducibility in pre-clinical animal
studies~\cite{Sanchez:2021}, biomedical sciences~\cite{Williams:2017,Montgomery:2019}, or
pattern recognition~\cite{Kerautret:2021}, with limited applicability outside these fields.

The book by Stodden, Leisch and Peng~\cite{Stodden:2014} comes, despite being a collection of
articles, close to what we need in academic teaching: it seeks to augment general
advice on reproducibility engineering with concrete technical details and examples.
However, almost all of the recommended tools---with the notable exception of knitr, which
we also include in our recommendations---stem from scientific research. At the time of this writing,
they are no longer maintained (VisTrail~\cite{Callahan:2006}), fail to build
(Sumatra), or are no longer available, apart from historical archives like the wayback
machine (CDE, SOLE). Given
that the book was published in 2014, this
underlines our strategy of relying on industrial, long-term maintained
tools, as
academic tools tend to break once project funding ceases~\cite{Fomel:2015}.

\paragraph{Online courses.}
Several MOOC platforms offer courses on reproducibility engineering (\eg, \webref{https://www.coursera.org/learn/reproducible-research}{Coursera},
\webref{https://www.edx.org/course/principles-statistical-and-computational-tools-for}{EdX},
\webref{https://www.fun-mooc.fr/en/courses/reproducible-research-methodological-principles-transparent-scie/}{Inria}).
They cover topics related to
software engineering (such as literate programming via notebooks),
but originate from outside the SWE community, (\eg, computational
biology or biomathematics).
Our own online course (cf.~Sec.~\ref{sec:intro}) assumes the computer
science perspective.

\section{Summary, Experience and Outlook}

Reproducibility engineering prepares students towards industry
careers, where sustainable long-term maintenance is important.
It should also become an entry-level requirement for PhD candidates.

In teaching and evaluating the class, we have observed that these goals
were satisfied.
Difficulties arise when technical details subtly impact reproducibility
(\eg, different CPU architecture between VM and host, or host CPU details).
A solution was to often add additional packages and layers
instead of identifying root causes of non-reproducibility. Consequently,
we find that details may matter to a larger extent than in other aspects
of software engineering.

Finally, we believe the effort contributes towards
solving the reproducibility crisis. Computer science and software
engineering seem, in a pivotal function, predestined for this purpose.

\begin{small}
\paragraph{Acknowledgements}
The joint effort was partly funded by the \emph{Lehr\-inno\-vationspool 2.0/2021-2022}
at the University of Passau. The authors were partly funded by \emph{Deutsche For\-schungsgemein\-schaft} (DFG, German Research Foundation) grant \#385808805,
and BMBF grant number 13N15645.
\end{small}

\section*{Authors' Profiles}

\newcommand{\name}[1]{\textsc{#1}}
\begin{wrapfigure}[6]{l}[0pt]{2.6cm}
\vspace*{-1.45em}\includegraphics[width=3.2cm]{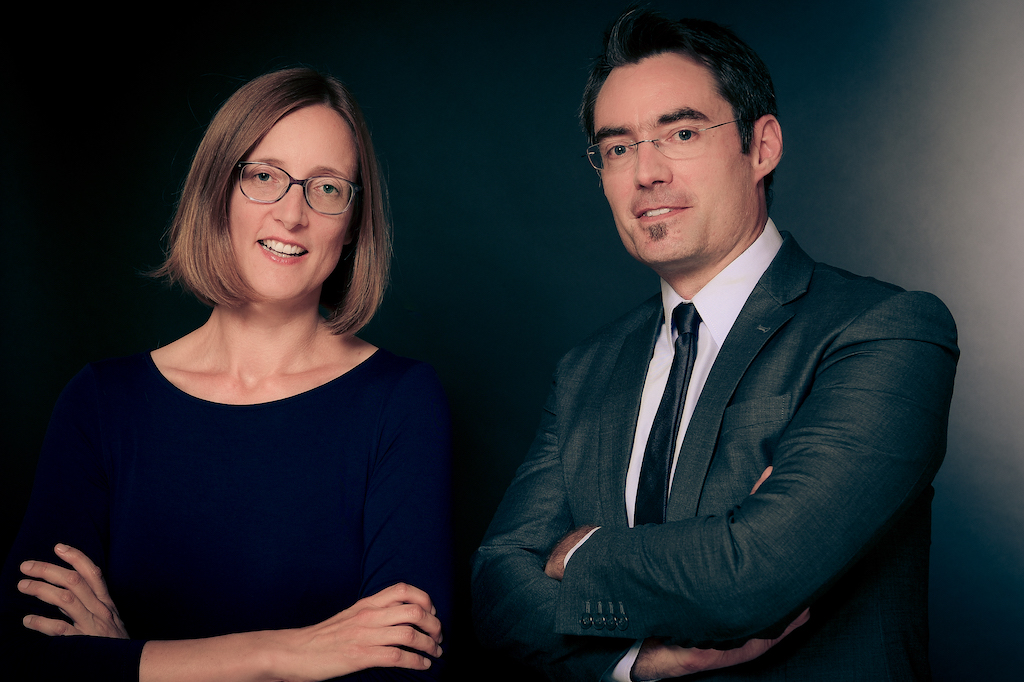}
\end{wrapfigure}
\name{Wolfgang Mauerer} (rhs) is a professor at Technical
University of Applied Sciences Regensburg, and a senior research scientist at Siemens
AG, Corporate Research. \name{Stefanie Scherzinger} (lhs) is a professor at  University of Passau, where she chairs the \emph{Scalable Database Systems} group. Together, they
have taught several
courses and tutorials~\cite{Mauerer:2020,Mauerer:2021}, including the inverted classroom on reproducibility engineering featured here. They have also
carried out reproduction studies of published research~\cite{Braininger:2020}.

\begin{wrapfigure}[3]{l}[0pt]{1.4cm}
\vspace*{-1.45em}\includegraphics[width=1.4cm]{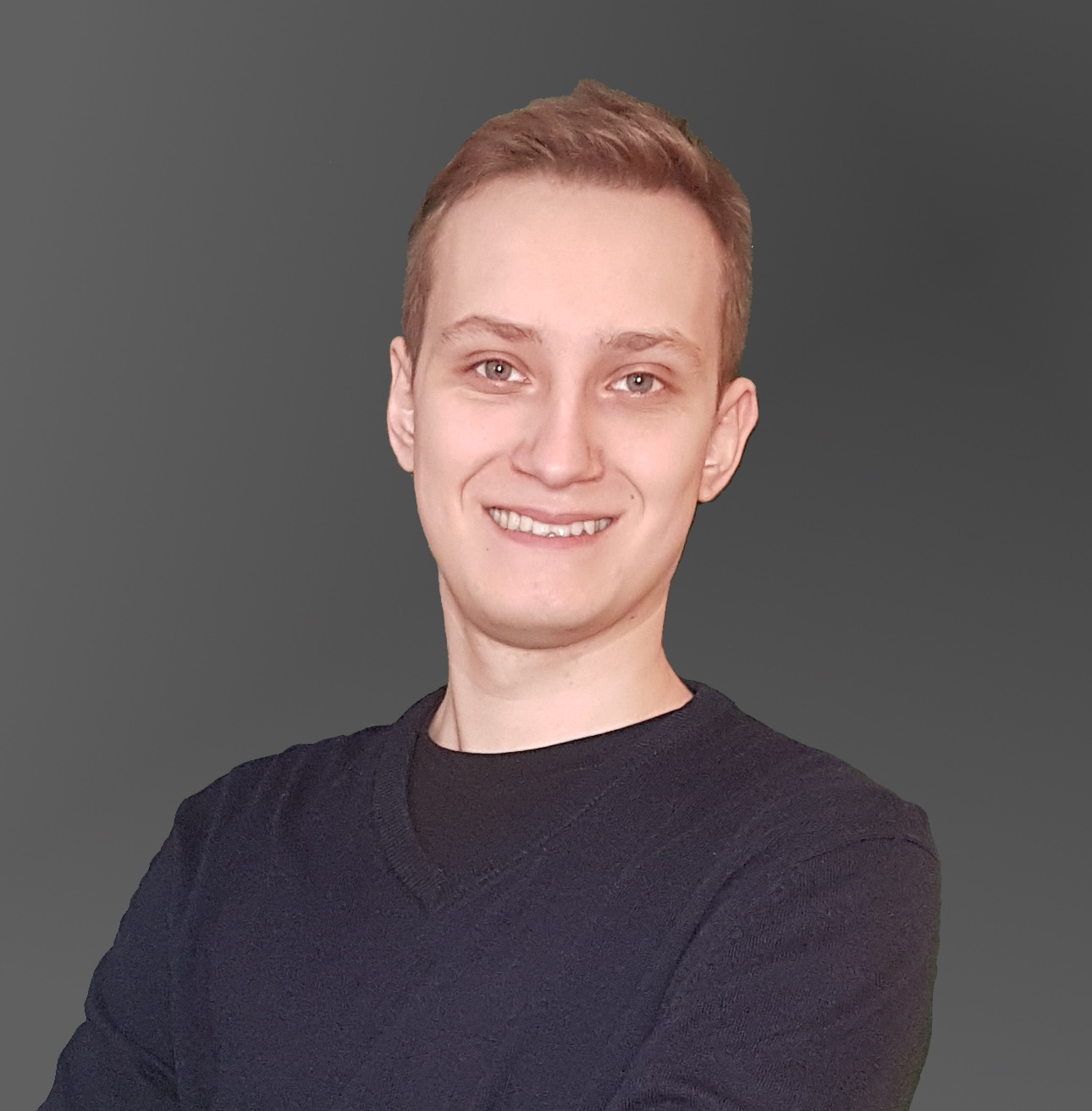}
\end{wrapfigure}
\noindent
\noindent\name{Stefan Klessinger}, M.~Sc., is a computer science researcher at University of Passau, and has tutored the joint RepEng course.

\bibliographystyle{ACM-Reference-Format}
\bibliography{literature.bib}

\end{document}